\documentclass[pre,aps,amsmath,amssymb,showpacs,preprint]{revtex4}
\usepackage{times}
\usepackage{epsf}
\usepackage{graphicx}
\usepackage{dcolumn}
\usepackage{bm}
\begin{document}


\title{Mathematical irrational numbers not so physically irrational}

\author{Y. J. Zhao}

\affiliation{Advanced Materials Laboratory, Fudan University,
Shanghai 200433, China}

\author{Y. H. Gao}

\affiliation{School of Machinery and Electricity, Henan Institute of
Science and Technology, Xinxiang 453003, China}

\author{J. P. Huang\footnote{Corresponding author. Tel: +86 21 55665227. Fax: +86 21 55665239. E-mail address: jphuang@fudan.edu.cn}}

\affiliation{Department of Physics and Surface Physics Laboratory
(National key laboratory), Fudan University, Shanghai 200433, China}

\date {\today}

\begin{abstract}

We investigate the topological structure of the decimal expansions
of the three famous naturally occurring irrational numbers, $\pi$,
$e$, and golden ratio, by explicitly calculating the diversity of
the pair distributions of the ten digits ranging from 0 to 9. And we
find that there is a universal two-phase behavior, which collapses
into a single curve with a power law phenomenon. We further reveal
that the two-phase behavior is closely related to general aspects of
phase transitions in physical systems. It is then numerically shown
that such characteristics originate from an intrinsic property of
genuine random distribution of the digits in decimal expansions.
Thus, mathematical irrational numbers are not  so {\it physically
irrational} as long as they have such an intrinsic property.

\end{abstract}





\maketitle

Diversity is an all-pervasive  characteristic of the world.
Literally, diversity in irrational numbers makes them so-called
``irrational".  In this work, we investigate irrational numbers
since numbers represent the reality. It is known that an
irrational number is a number which cannot be expressed as a
fraction of two integers, and that the decimal expansion of
irrational numbers never repeats or terminates, unlike rational
numbers. Here we focus on three naturally occurring irrational
numbers, namely, $\pi$, $e$, and golden ratio. The $\pi$ is the
ratio of a circle's circumference to its diameter in Euclidean
geometry. The $e$ is the unique irrational number such that the
value of the derivative of $f(x)=e^x$ at the point $x=0$ is
exactly 1. Two quantities are seen to be in the golden ratio if
the ratio between the sum of the two quantities and the larger one
is the same as the ratio between the larger one and the smaller.
The golden ratio is expressed as $(1+\sqrt{5})/2.$ The three
naturally occurring irrational numbers have abundant of uses in
physics, mathematics, and engineering. However,  our question is
if there is common interesting physical senses behind the three
naturally occurring irrational numbers. Our answer is true, as to
be addressed in this work.

Pair distributions are widely used, both experimentally and
theoretically, to treat physical problems, as a major descriptor for
various
microstructures~\cite{JuhasN06,MichelS07,JeongPRL05,HuangPRE05,PicklEPL08},
in which a pair distribution function describes the density of
inter-atomic or inter-particulate distances. We shall investigate
pair distributions of the ten digits (namely, 0, 1, $\cdots$, 9) in
the decimal expansions of the three irrational numbers, with a focus
on their diversity, so that one could get insight into the
connection between statistical physics and topological structures of
numbers.

The information entropy of a set of pair distributions provides a
measure of the diversity of the set. The greater the diversity of
the pair distributions in a set, the greater the entropy, while a
set with regular patterns has a small value for the entropy. In
general, information entropy has the following
characters~\cite{ShannonBSTJ48-1}: (i) Changing the value of one
of the probabilities by a very small amount should only change the
entropy by a small amount (continuity); (ii) The result should
keep unchanged if the probabilities are re-ordered (symmetry);
(iii) If all the probabilities are equally likely, information
entropy should be maximal (maximum); (iv) Adding or removing an
event with probability zero does not contribute to the entropy;
and (v) The amount of entropy should be the same independently of
how the process is regarded as being divided into parts
(additivity). Here (v) characterizes the information entropy of a
system with sub-systems, and it demands that the information
entropy of a system can be calculated from the information entropy
of its sub-systems if we know how the sub-systems interact with
each other.

For the first $10^4$ digits in the decimal expansions of the three
irrational numbers, $\pi$, $e$ and golden ratio, we calculate
their information entropy $H(N)$ of the pair distributions
according to
$$H(N) = -\sum_{i=0}^9\sum_j\left ( p_{i,j}\log p_{i,j}\right ),$$
with $p_{i,j} = N_{i,j}/\sum_jN_{i,j},$ where  $N_{i,j}$ denotes
the number of pairs with step size  $j$ (which is the distance
between every pair of two $i'$s)  for the digit $i$ ranging from 0
to 9.

\begin{figure}[h]
\includegraphics[width=400pt]{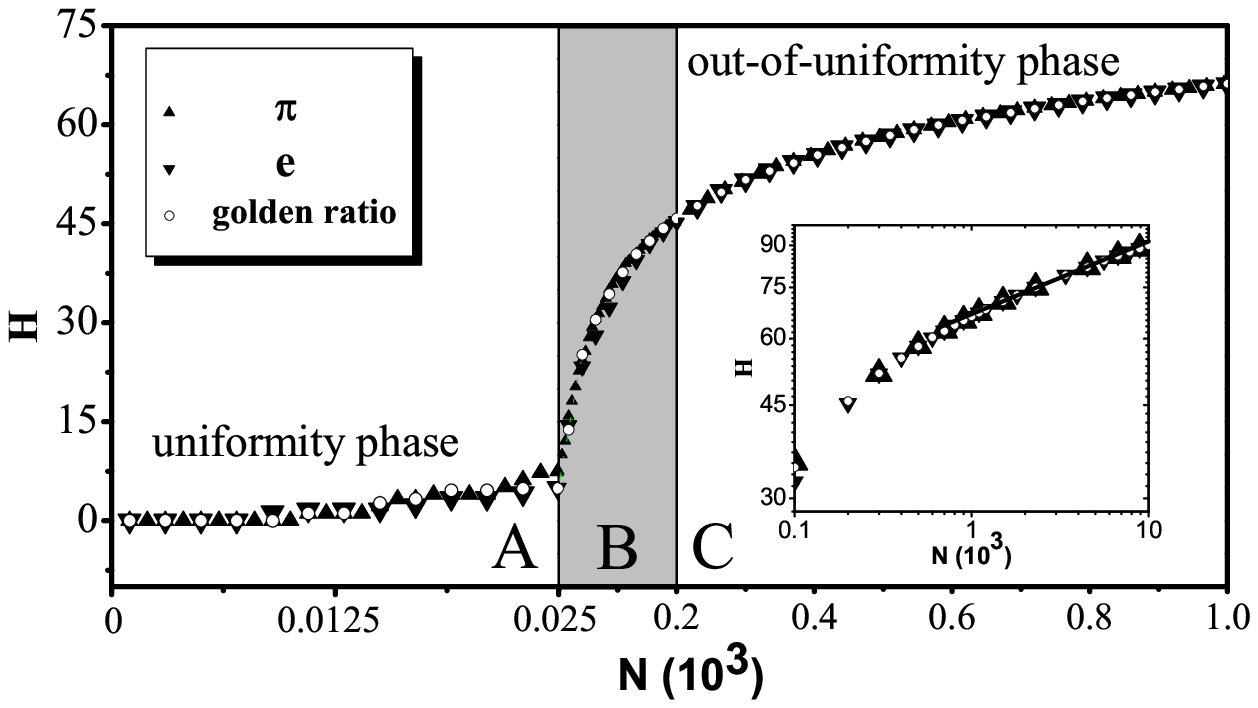}
\caption{}\label{fig1}
\end{figure}

Figure~\ref{fig1} displays the phase diagram for the existence of
two distinct phases for  the three naturally occurring irrational
numbers, $\pi$, $e$, and golden ratio, by plotting order parameter
$H$ (information entropy of the pair distributions of the ten
digits, ranging from 0 to 9) as a function of the number $N$ of the
first digits in the decimal expansions. The inset of Fig.~\ref{fig1}
shows a log-log plot of information entropy $H$ versus $N$ for the
same three irrational numbers, and a power-law relation appears, as
indicated by a straight line.  Such a single-curve collapse for the
three irrational numbers suggests that they are governed by the same
rule due to the same degree of disorder, which is, on the other
hand, implied by Fig.~\ref{fig2}. In fact, the two-phase behavior
shown for the ten digits holds for each digit as well. In this work,
we focus on the two-phase behavior for the ten digits as a whole.

We analyze the diversity of the pair distributions of the digits
ranging from 0 to 9, which is quantitatively expressed by their
information entropy $H$~\cite{ShannonBSTJ48-1}, and discover not
only a single-curve collapse but also the surprising existence of
a sudden change region (Region B), as shown in Fig.~\ref{fig1}.
For
 $N$ within Region
A, the value of $H$ is roughly zero; we interpret this as a
uniformity phase in which the diversity of the pair distributions
of the digits does not predominate. For Region C, $H$ increases
significantly; we interpret this as an out-of-uniformity phase in
which the diversity dominates and power-law phenomenon appears.

While phase transitions have been demonstrated to exist in various
mathematical systems~\cite{HoffmanCMP05}, our findings for the
irrational-number problem are identical to what is known to occur
in all phase-transition phenomena in physical systems. The
distinguishing characteristic of a phase transition is an abrupt
sudden change in one or more physical properties at a critical
threshold $\kappa_c$ of some control parameter $\kappa$. The
change in behavior at $\kappa_c$ can be quantified by an order
parameter $\phi(\kappa)$. For the irrational-number problem, we
find that the order parameter $\phi(\kappa)$ is given by the
information entropy $H(N)$. Region B bridges Regions A and C, and
denotes a sudden change region, indicating a sudden increase of
the diversity of the pair distributions. Next, we interpret these
two irrational-number phases according to Regions A and C.

In Region A, the entropy is roughly zero; we interpret this to be
the irrational-number uniformity phase, because the diversity of
the pair distributions of the ten digits does not predominate. In
the uniformity phase, there is almost no diversity, and
information entropies fluctuate around their {\it uniformity}
values, suggesting that most of the pair distributions donot have
diversity, but behave as uniformity (or none).

In Region C,  the entropy is large and increases step by step. We
interpret this to be the out-of-uniformity phase, because increasing
entropy suggests increasing diversity. So, in the out-of-uniformity
phase, the prevalent {\it uniformity} has changed, and the diversity
is now being driven to  new degree, which is consistent with the
fact that more pair distributions of the ten digits come to appear
with longer step sizes as $N$ increases. A power law is any
polynomial relationship that exhibits the property of scale
invariance, and has been explicitly used to characterize a
staggering number of natural
patterns~\cite{VandewallePRL01,ViswanathanN96,KeittN98}. The
observation of a power-law relation in the data for $\pi$, $e$, and
golden ratio, points to a specific kind of mechanisms that underly
the irrational-number phenomenon in question, and indicates a deep
connection between the irrational numbers. Such a connection
originates from the same degree of disorder, as shown in
Fig.~\ref{fig2}. Figure~\ref{fig2} presents  the occurrence
percentage of the distance between two consecutive digits in the
decimal expansions of the three irrational numbers ($\pi$, $e$, and
golden ratio) and a set of genuine random numbers. (Here ``genuine
random numbers'' mean those random numbers or quasi-random numbers
that have a very high degree of randomness. Throughout this work,
the phrase ``random numbers'' is simply used to indicate ``genuine
random numbers''. Similarly, in this work ``genuine random
distribution'' is used to represent the distribution with a very
high degree of randomness.) These random numbers were generated by
the commercial software Mathematica. The symbol lines for the three
irrational numbers are almost overlapped, which are further nearly
overlapped with that of the set of random numbers. This shows that
the three irrational numbers possess the same degree of disorder as
a group of random numbers. In other words, the above-mentioned
characteristics related to the topological structure of the decimal
expansions of $\pi$, $e$, and golden ratio actually originate from
an intrinsic property of genuine random distribution of the digits
in the decimal expansions.



Our results  suggest that there is a link between a mathematical
system with many digits (the irrational number) and the ubiquitous
phenomenon of phase transitions that occur in physical systems with
many units.

We hope that our work will stimulate further studies of number
physics. Here we have revealed a universal two-phase behavior for
the three famous naturally occurring irrational numbers, $\pi$, $e$,
and golden ratio. We should claim that the unique results obtained
herein for the three irrational numbers (as shown in
Fig.~\ref{fig1}) also hold for many other irrational numbers like
$\sqrt{2}$ and $\sqrt{3}$. It is also worth mentioning that the
results obtained from Fig.~\ref{fig1} (e.g., the single-curve
collapse) do not work for some other irrational numbers, e.g.,
$\sqrt{2.1}$. So far, there is no efficient method to sort such
irrational numbers clearly, except for investigating their decimal
expansions one by one. Nevertheless, we could safely conclude that
our findings are  a universal behavior for numerous irrational
numbers as long as they have an intrinsic property of genuine random
distribution of the digits in decimal expansions, and raise the
possibility that the topological structure of irrational numbers are
related to general aspects of phase transitions in physical systems.
In this sense, such mathematical irrational numbers seem to be not
so {\it physically irrational}.


\clearpage
\newpage

\section*{Figure Captions}

\begin{figure}[h]
\includegraphics[width=400pt]{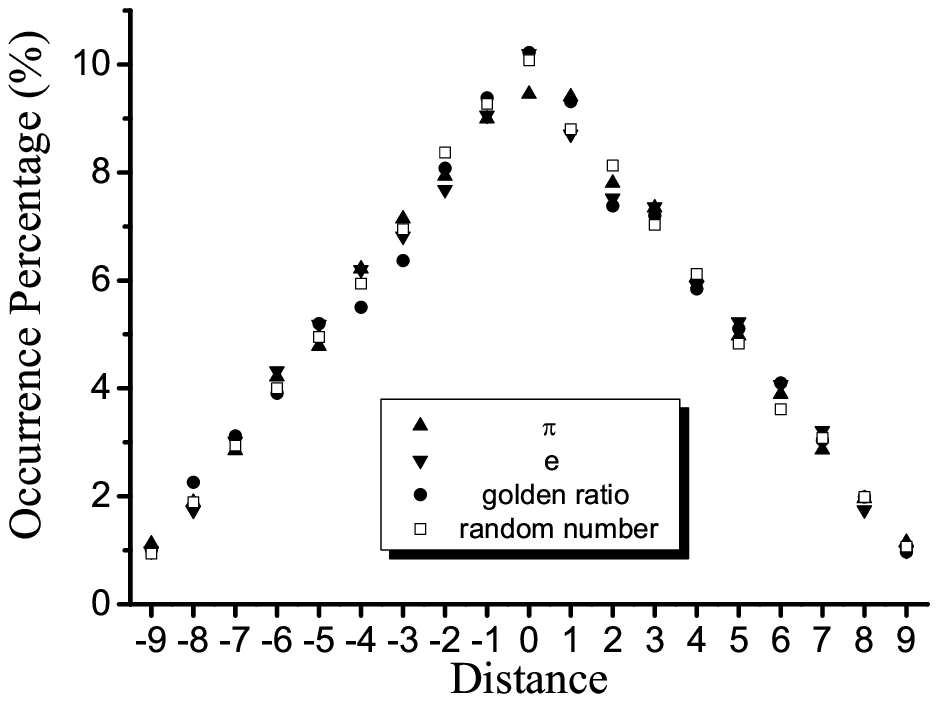}
\caption{}\label{fig2}
\end{figure}

Fig.~1. The  diagram representing  the existence of two distinct
phases for  the three naturally occurring irrational numbers, $\pi$,
$e$, and golden ratio, by plotting order parameter $H$ (information
entropy of the pair distributions of the ten digits, 0, 1, 2, 3, 4,
5, 6, 7, 8, and 9) as a function of the number $N$ of the first
digits in the decimal expansions. Region A: A uniformity phase;
Region B: An abrupt sudden change region; Region C: An
out-of-uniformity phase. Inset: Log-log plot of information entropy
$H$ versus $N$ for the same three irrational numbers, and a
power-law relation is shown as indicated by the straight line. The
symbol lines for the three irrational numbers are almost overlapped,
as investigated for $N$ up to $10^4.$ Such a single-curve collapse
suggests that they are governed by the same rule, due to the same
degree of disorder as presented by Fig.~\ref{fig2}.

Fig.~2.  The occurrence percentage of the distance between two
consecutive digits in the decimal expansions of the three irrational
numbers ($\pi$, $e$, and golden ratio) and a set of genuine random
numbers (which were generated by Mathematica), for $N=10^4$. The
four symbol lines are almost overlapped. Such a single-curve
collapse suggests that the three irrational numbers possess the same
degree of disorder as the set of  random numbers.

\clearpage
\newpage

\clearpage
\newpage

\end{document}